\documentclass[12pt,english,floatfix,showkeys,superscriptaddress,aps,prd,preprint]{revtex4}
\usepackage[latin1]{inputenc}
\usepackage[T1]{fontenc}
\usepackage{lmodern}
\usepackage{amsmath}
\usepackage{amssymb}
\usepackage{graphicx}
\usepackage{float}
\usepackage{esint}
\usepackage{dcolumn}
\usepackage{babel}
\usepackage{csquotes}
\usepackage{color}
\usepackage{caption}
\usepackage{subcaption}
\usepackage{slashed}
\usepackage{simplewick}
\usepackage{tikz-feynman}
\usepackage[]{tikz-feynman}
\usepackage{amsmath,latexsym}
\usepackage{ragged2e}
\usepackage{cancel}
\usepackage{orcidlink} % Para incluir ORCID

\usepackage{hyperref}
\hypersetup{
    colorlinks,
    citecolor=blue,
    filecolor=green,
    linkcolor=purple,
    urlcolor=red,
}

\usepackage{slashed}

\usepackage{hyperref}
\hypersetup{colorlinks,breaklinks,
			citecolor=[rgb]{0,0.0,1.0},
            urlcolor=[rgb]{0.0,0.0,1.0},
            linkcolor=[rgb]{0,0.5,0.9}}

\begin{document}
\title{Rarita-Schwinger model in Very Special Relativity}
%%%%%%%%%%%%%%%%%%%%%%%%%%%%%%%%%%%%%%%%%%%%%%%%%%%%%%%%%%%%%%%%%%%%%%
\author{M. C. Ara\'{u}jo}
\email{michelangelo@fisica.ufc.br}
\affiliation{Universidade Federal do Cariri, Av.~Tenente Raimundo Rocha, Cidade Universit\'{a}ria, 63048-080, Juazeiro do Norte, Cear\'{a}, Brazil}
%%%%%%%%%%%%%%%%%%%%%%%%%%%%%%%%%%%%%%%%%%%%%%%%%%%%%%%%%%%%%%%%%%%%%%%
\author{J. Furtado}
\email{job.furtado@ufca.edu.br}
\affiliation{Universidade Federal do Cariri, Av.~Tenente Raimundo Rocha, Cidade Universit\'{a}ria, 63048-080, Juazeiro do Norte, Cear\'{a}, Brazil}
%%%%%%%%%%%%%%%%%%%%%%%%%%%%%%%%%%%%%%%%%%%%%%%%%%%%%%%%%%%%%%%%%%%%%%%
\author{J. G. Lima \orcidlink{0000-0002-2443-0043}}
\email{junior.lima@uaf.ufcg.edu.br}
\affiliation{Departamento de F\'{\i}sica, Universidade Federal de Campina Grande, Caixa Postal 10071, 58429-900, Campina Grande, Para\'{\i}ba, Brazil}
%%%%%%%%%%%%%%%%%%%%%%%%%%%%%%%%%%%%%%%%%%%%%%%%%%%%%%%%%%%%%%%%%%%%%%%
\author{T. Mariz}
\email{tmariz@fis.ufal.br}
\affiliation{Instituto de F\'{i}sica, Universidade Federal de Alagoas, 57072-900, Macei\'{o}, Alagoas, Brazil}
%%%%%%%%%%%%%%%%%%%%%%%%%%%%%%%%%%%%%%%%%%%%%%%%%%%%%%%%%%%%%%%%%%%%%%%

\date{\today}

\begin{abstract}
In this work, we investigate vacuum polarization in the Rarita-Schwinger model within the framework of Very Special Relativity. We examine both massive and massless spin-$3/2$ fields coupled to the Maxwell field. The Mandelstam-Leibbrandt prescription is applied in order to evaluate the one-loop integrals, and we work within the $SIM(2)$ limit. 
\end{abstract}

%\pacs{04.70.-s,04.50.Kd,11.30.Cp,04.60.-m}

\maketitle

%%%%%%%%%%%%%%%%%%%%%%%%%%%%%%%%%%%%%%%%%%%%%%%%%%%%%%%%%%%%%%%%%%%%%%%%%%%%%%%%%%%%%%
\section{Introduction}\label{intro}

Although no definitive deviation from exact Lorentz invariance has been detected, there has recently been renewed interest in the possibility of small violations \cite{Kostelecky:1988zi,Carroll:1989vb,Colladay:1996iz,Coleman:1998ti,Colladay:1998fq,Stecker:2001vb,Carroll:2001ws,Kostelecky:2003fs,Kostelecky:2006ta,Kostelecky:2007zz}, motivated by the fact that such effects are conceivable in theories that aim to unify quantum mechanics and gravity \cite{Kostelecky:1988zi,Kostelecky:1991ak,Gambini:1998it,Bojowald:2004bb,Horava:2009uw,Cognola:2016gjy,Carroll:2001ws}. In general, three possible scenarios for the violation of Lorentz symmetry have been investigated. The first involves introducing non-dynamical (so-called spurion) tensor fields into the Lagrangian, which single out a preferred direction in space-time and thereby break the symmetry \cite{Coleman:1998ti,Coleman:1997xq,Myers:2003fd,Andrianov:2001zj,Alfaro:2006dd,Alfaro:2009mr}. The second arises from a mechanism of spontaneous Lorentz-symmetry breaking, in which non-dynamical (again spurion) tensor fields emerge as vacuum expectation values of fields belonging to some more fundamental theory \cite{Colladay:1996iz,Colladay:1998fq}. The third scenario -- which will receive greater emphasis here -- is provided by the framework of Very Special Relativity (VSR) \cite{Cohen:2006ky}.

The theory of VSR offers a particularly intriguing proposal by questioning whether the exact symmetry group of nature is in fact the Poincar\'{e} group, rather than one of its proper subgroups. Clearly, in order to preserve energy-momentum conservation, one must consider only those subgroups that include space-time translations, so that a proper subgroup of the Poincar\'{e} group is obtained by combining a proper Lorentz subgroup with these translations \cite{Cohen:2006ky}. Among the various possibilities, we highlight here the four-parameter similitude group $SIM(2)$. A notable feature of this group is that it admits no invariant tensor fields beyond those already invariant under the full Lorentz group. In other words, spurions cannot access scenarios in which the symmetry group of nature is $SIM(2)$ \cite{Cohen:2006ky}. Consequently, all classical effects of special relativity remain valid within VSR \cite{Cohen:2006ky,Alfaro:2023qib,Alfaro:2025uby}. Another interesting aspect of the VSR is that $CPT$ symmetry follows directly from $SIM(2)$, provided that amplitudes satisfy appropriate analyticity properties \cite{Cohen:2006ky}. This is an important point, since no violation of the $CPT$ symmetry has been observed to date, and it remains a fundamental symmetry of nature.

Another important point to mention is that the mass of neutrinos acquires a natural origin within VSR, without requiring the introduction of new particles or interactions that may or may not violate lepton-number conservation \cite{Cohen:2006ky,Dunn:2006xk,Cohen:2006ir}. Indeed, massive neutrinos that preserve lepton number are possible in VSR and arise from the observation that spin-$1/2$ particles may satisfy a modified Dirac equation, invariant under $SIM(2)$, containing terms proportional to $n_{\mu}/(n\cdot\partial)$, where $n_{\mu} = (1,0,0,1)$ is a null vector that defines a preferred direction \cite{Cheon:2009zx}. In this sense, although the theory is no longer Lorentz invariant, it is characterized by nonlocal terms that remain invariant under $SIM(2)$.

Also noteworthy is that VSR allows for the existence of massive photons while preserving the gauge symmetry of conventional quantum electrodynamics (QED) \cite{Cheon:2009zx,Alfaro:2015fha,Alfaro:2013uva}, a crucial feature for ensuring the unitarity and renormalizability of the theory \cite{Schwartz:2014sze}. In particular, the inclusion of massive photons and neutrinos does not affect the renormalizability of the model because the additional nonlocal terms mentioned above vanish in the large-momentum limit, thereby guaranteeing the same ultraviolet behavior as in strictly Lorentz-invariant theories \cite{Alfaro:2015fha,Alfaro:2013uva,Alfaro:2019koq}.

VSR has been extended to a wide range of domains since it was first proposed. Here, we highlight developments involving supersymmetry~\cite{Cohen:2006sc,Vohanka:2011aa}, curved spacetime~\cite{Gibbons:2007iu,Muck:2008bd}, noncommutative geometry~\cite{Sheikh-Jabbari:2008ybm,Das:2010cn}, and modifications involving the cosmological constant~\cite{Alvarez:2008uy}. Furthermore, it has been explored in the contexts of dark matter~\cite{Ahluwalia:2010zn}, cosmology~\cite{Chang:2013xwa}, Abelian gauge theories~\cite{Cheon:2009zx}, and non-Abelian gauge theories~\cite{Alfaro:2013uva}. Moreover, some observable consequences of VSR have been investigated in Refs.~\cite{Cohen:2006ir,Fan:2006nd,Dunn:2006xk}. However, as far as we are aware, particles of spin-$3/2$ have not yet been investigated in the context of VSR.

The objective of our work is therefore to investigate how vacuum polarization in a Rarita-Schwinger (RS) model is modified by the presence of the VSR. Originally proposed to describe spin-$3/2$ fields, RS theory has attracted considerable phenomenological interest and has been applied in various contexts. Notable examples include its role in describing gravitinos within supergravity (SUGRA)~\cite{Freedman:1976py,Das:1976ct,Gates:1983nr}, studies of scattering processes involving spin-$3/2$ particles~\cite{Delgado-Acosta:2009ulg,Antoniadis:2022jjy,Araujo:2024bug}, the modeling of hadron resonances~\cite{deJong:1992wm,Pascalutsa:1999zz,Bernard:2003xf}, and research on Lorentz-violating scenarios~\cite{Gomes:2022btc,Gomes:2023qkj,Gomes:2024qya}.

In this work, we investigate vacuum polarization within VSR for two distinct cases of RS theory, depending on whether the vector-spinor field is massive or massless. This distinction is important because the massless RS theory possesses its own gauge symmetry, which requires the introduction of gauge-fixing terms in the Lagrangian. %These terms significantly modify the propagator, even though VSR contributes masslike terms that encode the theory’s nonlocality and its departure from Lorentz invariance.
We also emphasize that we employ the Mandelstam-Leibbrandt (ML) prescription to evaluate the one-loop integrals and work within the $SIM(2)$ limit~\cite{Mandelstam:1982cb,Leibbrandt:1983pj,Alfaro:2016pjw}. As discussed in Refs.~\cite{Alfaro:2023qib,Alfaro:2025uby}, this is essential to ensure both gauge invariance and $SIM(2)$ invariance of the theory. 

The paper is organized as follows. In Sec.~\ref{RS}, we discuss the Lagrangian of the RS field coupled to the Maxwell field in the VSR scenario, including its general form and the corresponding Feynman rules. In Sec.~\ref{VP}, we compute the one-loop contributions to the vacuum polarization for both the massive and massless cases. Finally, in Sec.~\ref{CO}, we summarize the main results of the paper. Throughout this work, we employ natural units and adopt the Minkowski metric $g^{\mu\nu} = \mathrm{diag}(1, -1, -1, -1)$.

\section{RS model in the VSR framework}\label{RS}

In this section, we are interested in studying the RS Lagrangian of spin-$3/2$ field in the framework of VSR. The Lagrangian density is given by
\begin{eqnarray}\label{densmodelgeneral}
\mathcal{L} = \Bar{\psi}^{\mu}\Lambda_{\mu \nu}\psi^{\nu},
\end{eqnarray}
where the operator $\Lambda_{\mu\nu}$ can be written as
\begin{equation} \label{operatorL}
\Lambda^{\mu\nu} = \frac{i}{2}\left\{\sigma^{\mu\nu}, \left(i\slashed{\mathcal{D}} -m +\frac{m_\nu^{2}}{2}\frac{i\slashed{n}}{n\cdot \mathcal{D}} \right)\right\}.
\end{equation}
Here, $\sigma^{\mu\nu}=\frac{i}{2}[\gamma^{\mu},\gamma^{\nu}]$ denotes the Dirac sigma matrix, $\mathcal{D}_{\mu}=\partial_{\mu}-ieA_{\mu}$ is the covariant derivative, $m$ represents the usual fermion mass, $m_\nu$ is a mass parameter assigned by the VSR, and $n^{\mu}$ is a fixed null vector satisfying $n\cdot n=0$. The above Lagrangian can be equivalently rewritten in an expanded form as
\begin{eqnarray}\label{RSL}
\mathcal{L} &=& \bar{\psi}_{\mu}((i\tilde{\slashed{\mathcal{D}}} -m)g^{\mu\nu} -i(\gamma^{\mu} \tilde{\mathcal{D}}^{\nu} +\gamma^{\nu}\tilde{\mathcal{D}}^{\mu}) +i\gamma ^{\mu }\tilde{\slashed{\mathcal{D}}}\gamma^{\nu} +m\gamma^{\mu}\gamma^{\nu})\psi_{\nu},       
\end{eqnarray}
where
\begin{equation}
\tilde{\mathcal{D}}_\mu = \mathcal{D}_\mu + \frac{m_\nu^{2}}{2}\frac{n_\mu}{n\cdot \mathcal{D}}.
\end{equation}

Now, within the VSR framework, the Feynman rules yield the propagator of the RS model in $D$ dimensions as
\begin{equation}\label{Gm}
iG^{\mu\nu}(p)=i \frac{\slashed{\tilde{p}}+m}{\tilde{p}^2-m^2}\left(g^{\mu\nu}-\frac{1}{D-1}\gamma^{\mu}\gamma^{\nu}-\frac{D-2}{D-1}\frac{\tilde{p}^{\mu}\tilde{p}^{\nu}}{m^2}-\frac{1}{D-1}\frac{\gamma^{\mu}\tilde{p}^{\nu}-\gamma^{\nu}\tilde{p}^{\mu}}{m}\right),
\end{equation}
while the interaction vertices of interest are given by
\begin{eqnarray}
-ieV^{\lambda\mu\nu}(p,p')=-ie\left({g_{\alpha}}^{\lambda}+\frac{m_\nu^2}{2}\frac{n_\alpha n^\lambda}{(n\cdot p)(n\cdot p')}\right)\gamma^{\mu\alpha\nu},
\end{eqnarray}
and
\begin{eqnarray}
-ieV^{\kappa\lambda\mu\nu}(p,p',k,k')=-ie\frac{m_\nu^2}{2}\frac{n_\alpha n^\kappa n^\lambda}{(n\cdot p)(n\cdot p')}\left(\frac{1}{n\cdot(p+k)}+\frac{1}{n\cdot(p+k')}\right)\gamma^{\mu\alpha\nu},
\end{eqnarray}
with $\tilde{p}_\mu=p_\mu-\frac{m_\nu^2}{2}\frac{n_\mu}{n\cdot p}$ and $\gamma^{\mu\alpha\nu}=g^{\mu\nu}\gamma^\alpha -g^{\mu\alpha}\gamma^\nu -g^{\nu\alpha}\gamma^\mu +\gamma^\mu\gamma^\alpha\gamma^\nu$. It is worth mentioning that the propagator in Eq. \eqref{Gm} is not applicable when the fermion mass $m$ vanishes. In this limit, the RS theory possesses an intrinsic gauge symmetry that is not broken by the extra mass term induced by VSR. Therefore, for $m=0$, the appropriate propagator is \begin{equation}\label{Gms}
i G^{\mu\nu}(p)=\frac{i}{\tilde{p}^2}\left(\slashed{\tilde{p}}g^{\mu\nu}-\frac{2}{D-2}(\gamma^{\mu}\tilde{p}^{\nu}+\gamma^{\nu}\tilde{p}^{\mu})+\frac{1}{D-2}\gamma^\mu\slashed{\tilde{p}}\gamma^\nu\right) +\frac{i}{\tilde{p}^4}\left(\frac{4}{D-2}-\frac{1}{\xi}\right)\tilde{p}^\mu\slashed{\tilde{p}}\tilde{p}^\nu,
\end{equation}
which is obtained by supplementing the Lagrangian with a gauge-fixing term of the form ${\cal L}_{GF}=-\xi\bar{\psi}_{\mu}\gamma^\mu i\slashed{\partial}\gamma^\nu\psi_{\nu}$. It should also be noted that, with this choice, the vertices remain unchanged.

\section{Vacuum polarization}\label{VP}

\begin{figure}[h!]
\centering
\begin{tabular}{cc}
\raisebox{10mm}{\includegraphics[scale=0.5]{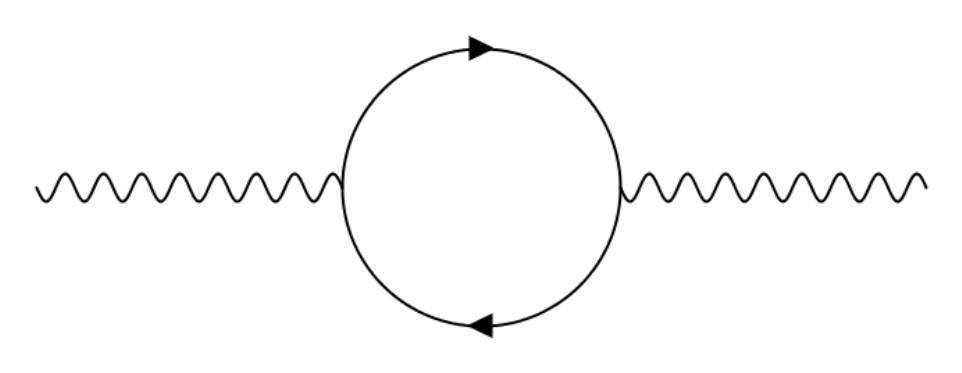}} &
\includegraphics[scale=0.5]{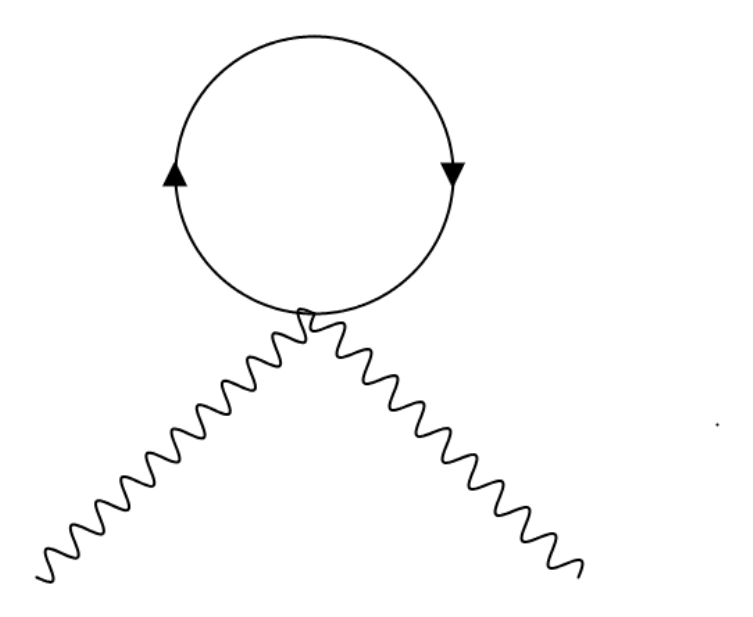} \\
\end{tabular}
\caption{Vacuum polarization one-loop graphs}
\label{Fig1}
\end{figure}

As an initial investigation of the RS model in the VSR framework, we evaluate the one-loop vacuum polarization for both massive and massless cases. The corresponding Feynman diagrams are presented in Fig.~(\ref{Fig1}). 

\subsection{Massive case}\label{MV}

In this subsection, we compute the photon self-energy arising from the one-loop vacuum polarization diagrams involving the propagator in Eq.~\eqref{Gm}. The two contributions in question are represented by
\begin{eqnarray}
\Pi_1^{\mu\nu} &=& ie^2\,{\rm tr}\int\frac{d^4p}{(2\pi)^4}
G_{\alpha\beta}(p)V^{\mu\beta\gamma}(p,p_1)G_{\gamma\delta}(p_1)V^{\nu\delta\alpha}(p_1,p),
\end{eqnarray}
and
\begin{eqnarray}
\Pi_2^{\mu\nu} &=& -ie^2\,{\rm tr}\int\frac{d^4p}{(2\pi)^4}
G_{\alpha\beta}(p)V^{\mu\nu\beta\alpha}(p,p,k,-k),
\end{eqnarray}
where $p_1=p+k$, with $k^\mu$ denoting the momentum of the external photon, and the trace is evaluated over the Dirac matrices.

After performing the trace, we obtain 
\begin{eqnarray}\label{Pi1}
\Pi_1^{\mu\nu} &=& ie^2\int\frac{d^4p}{(2\pi)^4}\frac{1}{(p^2-m_\nu^2-m^2)(p_1^2-m_\nu^2-m^2)} \nonumber\\
&\times&\left[
N_1^{\mu\nu}\frac{(n\cdot p_1)^{3}}{(n\cdot p)^{3}}+
N_2^{\mu\nu}\frac{(n\cdot p_1)^{2}}{(n\cdot p)^{2}}+
N_3^{\mu\nu}\frac{(n\cdot p_1)^{2}}{(n\cdot p)^{3}}+
N_4^{\mu\nu}\frac{n\cdot p_1}{n\cdot p}+
N_5^{\mu\nu}\frac{n\cdot p_1}{(n\cdot p)^{2}}
\right.\nonumber\\ &+&\left.
N_6^{\mu\nu}\frac{n\cdot p_1}{(n\cdot p)^{3}}+
N_7^{\mu\nu}\frac{n\cdot p}{n\cdot p_1}+
\frac{N_8^{\mu\nu}}{(n\cdot p_1)(n\cdot p)}+
N_9^{\mu\nu}\frac{(n\cdot p)^{2}}{(n\cdot p_1)^{2}}+
N_{10}^{\mu\nu}\frac{n\cdot p}{(n\cdot p_1)^{2}} \right.\nonumber\\ &+&\left.
N_{11}^{\mu\nu}\frac{(n\cdot p)^{3}}{(n\cdot p_1)^{3}}+
N_{12}^{\mu\nu}\frac{(n\cdot p)^{2}}{(n\cdot p_1)^{3}}+
N_{13}^{\mu\nu}\frac{n\cdot p}{(n\cdot p_1)^{3}}+
\frac{N_{14}^{\mu\nu}}{n\cdot p}+
\frac{N_{15}^{\mu\nu}}{(n\cdot p)^{2}}
\right.\nonumber\\ &+&\left.
\frac{N_{16}^{\mu\nu}}{n\cdot p_1}+
\frac{N_{17}^{\mu\nu}}{(n\cdot p_1)^{2}}+
N_{18}^{\mu\nu}\right]
\end{eqnarray}
and
\begin{eqnarray}\label{Pi2}
\Pi_2^{\mu\nu} &=& -ie^2\int\frac{d^4p}{(2\pi)^4}
\frac{4(2-D)m_\nu^2}{p^2-m_\nu^2-m^2}\frac{n^\mu n^\nu}{(n\cdot p_1)(n\cdot p_2)},
\end{eqnarray}
where $p_2=p-k$ and $N_i = N_i(p,k)$ ($i = 1,2,\dots,18$) are functions that do not depend on the product $n \cdot p$. 

We can further simplify Eq.~\eqref{Pi1}, since within the framework of $SIM(2)$-invariant regularization (see Refs.~\cite{Alfaro:2023qib,Alfaro:2025uby}), any integral involving negative powers of $n \cdot p$ vanishes. That is, 
\begin{eqnarray}
\int\frac{d^Dp}{(2\pi)^D}\frac{1}{(p^2+2p\cdot q-m^2)^a}\frac{1}{(n\cdot p)^b}=0,
\end{eqnarray}
and 
\begin{eqnarray}
\int\frac{d^Dp}{(2\pi)^D}\frac{p_{\alpha_1}\cdots p_{\alpha_n}}{(p^2+2p\cdot q-m^2)^a}\frac{1}{(n\cdot p)^b}=0,
\end{eqnarray}
with the latter being obtained from the former by taking derivatives with respect to $q^{\mu}$, with $b > 0$. Thus, after applying Feynman parametrization and using the expressions
\begin{eqnarray}
\frac{n\cdot p_1}{n\cdot p} &=& 1+\frac{n\cdot k}{n\cdot p}, \\
\frac{n\cdot p}{n\cdot p_1} &=& 1-\frac{n\cdot k}{n\cdot p_1}, \\
\frac{1}{(n\cdot p)(n\cdot p_1)} &=& \frac{1}{n\cdot k}\left(\frac{1}{n\cdot p}-\frac{1}{n\cdot p_1}\right), \\
\frac{1}{(n\cdot p_1)(n\cdot p_2)} &=& \frac{1}{2(n\cdot k)}\left(\frac{1}{n\cdot p_2}-\frac{1}{n\cdot p_1}\right),
\end{eqnarray}
we can conclude that all integrals corresponding to the coefficients $N_{3}^{\mu\nu}$, $N_{5}^{\mu\nu}$, $N_{6}^{\mu\nu}$, $N_{8}^{\mu\nu}$, $N_{10}^{\mu\nu}$, $N_{12}^{\mu\nu}$, $N_{13}^{\mu\nu}$, $N_{14}^{\mu\nu}$, $N_{15}^{\mu\nu}$, $N_{16}^{\mu\nu}$, and $N_{17}^{\mu\nu}$ vanish, allowing us to write
\begin{eqnarray}\label{Pi1a}
\Pi_1^{\mu\nu} &=& ie^2\int_0^1dx\ \mu^{4-D}\int\frac{d^Dp}{(2\pi)^D}\frac{1}{(p^2+2p\cdot k(1-x)+k^2(1-x)-\tilde{m}^2)^2} \nonumber\\
&\times& \left(N_1^{\mu\nu}+N_2^{\mu\nu}+N_4^{\mu\nu}+N_7^{\mu\nu}+N_9^{\mu\nu}+N_{11}^{\mu\nu}+N_{18}^{\mu\nu}\right),
\end{eqnarray}
where the explicit form of the remaining coefficients are 
\begin{eqnarray}
N_1^{\mu\nu} = N_{11}^{\mu\nu} = \frac{(D-2)^2 m_{\nu }^6 g^{\mu  \nu }}{2 (D-1)^2 m^4},
\end{eqnarray}
\begin{eqnarray}
N_2^{\mu\nu} = N_9^{\mu\nu} &=& \frac{m_{\nu }^4}{(D-1)^2 m^4}(g^{\mu  \nu } (((D-2) D-4) m^2-3 (D-2)^2 (p\cdot p_1)) \nonumber\\
&+&(D-2)^2 (p^{\mu } p_1^{\nu }+p_1^{\mu } p^{\nu })),
\end{eqnarray}
\begin{eqnarray}
N_{4a}^{\mu\nu} = N_{7a}^{\mu\nu} &=& \frac{m_{\nu }^2 g^{\mu  \nu }}{2 (D-1)^2 m^4} (4 (D-2)^2 p^2 (m^2+m_{\nu }^2-p_1^2)-8 ((D-2) D-4) m^2 (p\cdot p_1) \nonumber\\
&+&(D-2) (4 ((D-4) D+1) m^4+(D-2) (4 p_1^2 (m^2+m_{\nu }^2)-m_{\nu }^2 \nonumber\\
&\times& (8 m^2+m_{\nu }^2))) +12 (D-2)^2 (p\cdot p_1)^2),
\end{eqnarray}
\begin{eqnarray}
N_{4b}^{\mu\nu} = N_{7b}^{\mu\nu} &=& -\frac{2 m_{\nu }^2}{(D-1)^2 m^4} (p^{\mu } ((D-2)^2 p^{\nu } (m^2+m_{\nu }^2-p_1^2)+2 p_1^{\nu } ((D-2)^2 (p\cdot p_1) \nonumber\\
&-&(D-4) m^2)) +p_1^{\mu } (2 p^{\nu } ((D-2)^2 (p\cdot p_1)-(D-4) m^2) \nonumber\\
&+&(D-2)^2 p_1^{\nu } (m^2+m_{\nu }^2-p^2))),
\end{eqnarray}
\begin{eqnarray}
N_{18a}^{\mu\nu} &=& \frac{2 m_{\nu }^2 g^{\mu  \nu }}{(D-1)^2 m^4} (-2 p^2 (((D-6) D+12) m^2+(D-2)^2 (p\cdot p_1))+m^2 (4 (D-2) D m^2 \nonumber\\
&+&((3 D-14) D+20) m_{\nu }^2-2 ((D-6) D+12) p_1^2)+(D-2)^2 (p\cdot p_1) \nonumber\\ 
&\times&(4 m^2-m_{\nu }^2-2 p_1^2)),
\end{eqnarray}
\begin{eqnarray}
N_{18b}^{\mu\nu} &=& \frac{4 g^{\mu  \nu }}{(D-1)^2 m^4} ((D-2) (D-1)^2 m^6-(D-2) m^2 p_1^2 (D m^2+(D-2) (p\cdot p_1)) \nonumber\\
&+&p^2 (-(D-2) D m^4+((D-6) D+12) m^2 p_1^2-(D-2)^2 (m^2-p_1^2) (p\cdot p_1)) \nonumber\\
&+&(p\cdot p_1) (-(D-2) ((D-4) D+1) m^4+((D-2) D-4) m^2 (p\cdot p_1) \nonumber\\
&-&(D-2)^2 (p\cdot p_1){}^2)),
\end{eqnarray}
\begin{eqnarray}
N_{18c}^{\mu\nu} &=& \frac{4}{(D-1)^2 m^4} (p_1^{\mu } p_1^{\nu } (m^2 ((D-2) D m^2+((D-6) D+12) m_{\nu }^2-((D-6) D+12) p^2) \nonumber\\
&+&(D-2)^2 (p\cdot p_1) (m^2+m_{\nu }^2-p^2))+p^{\mu } p^{\nu } ((D-2) D m^4+((D-6) D+12) m^2 \nonumber\\
&\times&(m_{\nu }^2-p_1^2)+(D-2)^2 (p\cdot p_1) (m^2+m_{\nu }^2-p_1^2))),
\end{eqnarray}
\begin{eqnarray}
N_{18d}^{\mu\nu} &=& \frac{2 (p_1^{\mu } p^{\nu }+p^{\mu } p_1^{\nu })}{(D-1)^2 m^4} ((D-2) (2 ((D-3) D+1) m^4+(D-2) m_{\nu }^4) \nonumber\\
&-&4 (D-4) m^2 (p\cdot p_1)+2 (D-2)^2 (p\cdot p_1)^2),
\end{eqnarray} 
with $N_{4}^{\mu\nu}=N_{4a}^{\mu\nu}+N_{4b}^{\mu\nu}$, $N_{7}^{\mu\nu}=N_{7a}^{\mu\nu}+N_{7b}^{\mu\nu}$, $N_{18}^{\mu\nu}=N_{18a}^{\mu\nu}+N_{18b}^{\mu\nu}+N_{18c}^{\mu\nu}+N_{18d}^{\mu\nu}$ and $\tilde{m}^2=m_\nu^2+m^2$. Note that dimensional regularization has been employed, which consists in extending spacetime from $4$ to $D$ dimensions, so that the integration measure is modified from $d^4p/(2\pi)^4$ to $\mu^{4-D} d^Dp/(2\pi)^D$, with $\mu$ acting as a mass regulator. Furthermore, the contribution in Eq.~\eqref{Pi2} vanishes identically.

Now, applying the shift $p_\mu \to p_\mu-k_\mu (1-x)$, we can rewrite Eq.~\eqref{Pi1a} as 
\begin{eqnarray}
\Pi_1^{\mu\nu}=\Pi_{1a}^{\mu\nu}+\Pi_{1b}^{\mu\nu}+\Pi_{1c}^{\mu\nu}+\Pi_{1d}^{\mu\nu},
\end{eqnarray}
with
\begin{eqnarray}
\Pi_{1a}^{\mu\nu} &=& ie^2\int_0^1dx\ \mu^{4-D}\int\frac{d^Dp}{(2\pi)^D}\frac{1}{(p^2-M^2)^2} \frac{4 g^{\mu  \nu }}{(D-1)^2 D m^4}
(k^2 (-(D-2) (D^2 m^4-D^2 M^4 \nonumber\\
&+&(D-1)^2 D m^4 x^2-(D-1)^2 D m^4 x+4 D m^2 M^2+4 D M^4-12 m^2 M^2-4 M^4) \nonumber\\
&+&2 m_{\nu }^2 ((D-4) D m^2-(D-2)^2 (D-1) M^2)+(D-2)^2 D m_{\nu }^4)-(D-2)^2 \nonumber\\
&\times& k^4 (x-1) x (D m^2+D m_{\nu }^2-(D-2) M^2)+(D-2) (D-1)^2 \nonumber\\
&\times& m^4 (D m^2+D m_{\nu }^2-(D-2) M^2)),
\end{eqnarray}
\begin{eqnarray}
\Pi_{1b}^{\mu\nu} &=& ie^2\int_0^1dx\ \mu^{4-D}\int\frac{d^Dp}{(2\pi)^D}\frac{1}{(p^2-M^2)^2} \frac{4 k^{\mu } k^{\nu }}{(D-1)^2 D m^4} ((D-2) (D^2 m^4-D^2 M^4 \nonumber\\
&+&2 (D-1)^2 D m^4 x^2-2 (D-1)^2 D m^4 x+4 D m^2 M^2+4 D M^4-12 m^2 M^2-4 M^4) \nonumber\\
&+&(D-2)^2 k^2 (x-1) x (D m^2+D m_{\nu }^2-(D-2) M^2)+2 m_{\nu }^2 ((D-2)^2 (D-1) M^2 \nonumber\\
&-&(D-4) D m^2)-(D-2)^2 D m_{\nu }^4),
\end{eqnarray}
\begin{eqnarray}
\Pi_{1c}^{\mu\nu} &=& -ie^2\int_0^1dx\ \mu^{4-D}\int\frac{d^Dp}{(2\pi)^D}\frac{1}{(p^2-M^2)} \frac{4 (D-2) g^{\mu  \nu }}{(D-1)^2 D m^4} (k^2 (-(D-2)^2 k^2 (x-1) x \nonumber\\
&+&4  m^2(D-3) +2 (D-2) (D-1) m_{\nu }^2-2 (D-2)^2 M^2)+(D-2) (D-1)^2 m^4),\hspace{0.5cm} 
\end{eqnarray}
\begin{eqnarray}
\Pi_{1d}^{\mu\nu} &=& -ie^2\int_0^1dx\ \mu^{4-D}\int\frac{d^Dp}{(2\pi)^D}\frac{1}{(p^2-M^2)} \frac{4 (D-2) k^{\mu } k^{\nu }}{(D-1)^2 D m^4} ((D-2)^2 k^2 (x-1) x \nonumber\\
&-&4 (D-3) m^2-2 (D-1) (D-2) m_{\nu }^2+2 (D-2)^2 M^2),
\end{eqnarray}
and $M^2=k^2(x-1)x+\tilde{m}^2$. Therefore, performing the momentum integrals, the total expression becomes
\begin{eqnarray}
\Pi_1^{\mu\nu} &=& -e^2\int_0^1dx\ \frac{2^{3-D} e^{-\frac{1}{2} i \pi  D} \pi ^{-\frac{D}{2}} \mu ^4 \left(\mu ^2\right)^{-\frac{D}{2}} \Gamma \left(2-\frac{D}{2}\right) (-M^2)^{\frac{D-4}{2}} }{(D-1)^2 D m^4}(k^2g^{\mu\nu}-k^\mu k^\nu) \nonumber\\
&&\times (k^2 (x-1) x ((((D-4) D+2) D+8) m^2+(D-4) (D-2) m_{\nu }^2)+(D-2)^2 \nonumber\\
&&\times (D+1) k^4 (x-1)^2 x^2+m^4 ((D-2) D (-((D-1)^2 (x-1) x)-2)+4) \nonumber\\
&&-2 (D-2) m_{\nu }^2 (2 m^2+m_{\nu }^2)).
\end{eqnarray} 
Finally, expanding the above expression around $D=4$ and perform the integration over $x$, we obtain
\begin{eqnarray}\label{Pi1d}
\Pi_1^{\mu\nu} &=& \frac{e^2}{54 \pi ^2 \epsilon' m^4} (4 k^2 m^2-k^4+6 (2 m^2 m_{\nu }^2+m_{\nu }^4)) (k^2 g^{\mu  \nu }-k^{\mu } k^{\nu }) 
+\frac{e^2}{108 \pi ^2 m^4} \nonumber\\
&&\times [(2 k^2 (m^2-2 m_{\nu }^2)+(25 m^4+26 m^2 m_{\nu }^2+5 m_{\nu }^4)-k^4-72 m^4 (m^2+m_{\nu }^2)k^{-2}) \nonumber\\
&&+(4 k^2 (m_{\nu }^2-m^2)-4 (m^4+4 m^2 m_{\nu }^2)+2k^4+72 m^4 (m^2+m_{\nu }^2)k^{-2}) \nonumber\\
&&\times \sqrt{4 (m^2+m_{\nu }^2)k^{-2}-1}\ \cot ^{-1}(\sqrt{4 (m^2+m_{\nu }^2)k^{-2}-1})] (k^2 g^{\mu  \nu }-k^{\mu } k^{\nu }),
\end{eqnarray}
where $\frac 1{\epsilon'}=\frac 1\epsilon-\ln\frac m{\mu'}$, with $\epsilon=4-D$ and $\mu'^2=4\pi\mu^2e^{-\gamma}$. It is straightforward to verify that this expression is gauge invariant and that the limit $m_\nu \to 0$ is smooth, in which case we recover the conventional result \cite{Barua:1978ck}. Furthermore, we observe that the divergent term in Eq.~\eqref{Pi1d}, that depends on $m_\nu$, involves only the transverse projector ($k^2g^{\mu\nu}-k^\mu k^\nu$), just as in QED. 

% $M^2=k^2(x-1)x+m_\nu^2+m^2$

\subsection{Massless case}\label{ML}

We next evaluate the photon self-energy in the massless case. By using the propagator in Eq.~\eqref{Gms} and performing the Dirac trace, we obtain an expression analogous to Eq.~\eqref{Pi1}, allowing us to write
\begin{eqnarray}\label{Pi1b}
\Pi_1^{\mu\nu} &=& ie^2\int_0^1dx\ \mu^{4-D}\int\frac{d^Dp}{(2\pi)^D}\frac{1}{(p^2+2p\cdot k(1-x)+k^2(1-x)-m_\nu^2)^4} \nonumber\\
&\times&(N_1^{\mu\nu}+N_2^{\mu\nu}+N_4^{\mu\nu}+N_7^{\mu\nu}+N_9^{\mu\nu}+N_{11}^{\mu\nu}+N_{18}^{\mu\nu}),
\end{eqnarray}
with
\begin{eqnarray}
N_{1}^{\mu\nu}=N_{11}^{\mu\nu}=\frac{(D -4\xi -2)^2 m^{6}_{\nu}g^{\mu  \nu }}{2(D -2)^2 \xi^2},
\end{eqnarray}
\begin{eqnarray}
N_{2}^{\mu\nu}=N_{9}^{\mu\nu}=\frac{(D-4 \xi -2)^2 m^{4}_{\nu}}{(D-2)^2 \xi ^2}(-3 (p\cdot p_1) g^{\mu  \nu }+p^{\mu } p_1^{\nu }+p_1^{\mu } p^{\nu }),
\end{eqnarray}
\begin{eqnarray}
N_{4a}^{\mu\nu} = N_{7a}^{\mu\nu} &=& \frac{m_{\nu}^2 g^{\mu\nu}}{2(D -2)^2\xi^2}((4(D((D-5)D +8) -8)\xi^2 +8(D- 2)\xi -(D -2)^2)m_{\nu}^4 \nonumber \\
&+& 4((D((D -5)D +8) -20)\xi^2 +8(D -2)\xi -(D -2)^2)(p^2 p_1^2 \nonumber \\
&-& (p^2 +p_1^2)m_{\nu}^2) +12 (D - 4\xi -2)^2(p\cdot p_1)^2),
\end{eqnarray}
\begin{eqnarray}
N_{4b}^{\mu\nu} = N_{7b}^{\mu\nu} &=& -\frac{(D -4\xi -2)^2 m_{\nu}^2} {(D - 2)^2\xi^2}((p^{\mu}p^{\nu}(m_{\nu}^2 - p_1^2) +2 p^{\mu}p_1^{\nu} (p\cdot p_1)) +p_1^{\mu}p_1^{\nu}(m_{\nu}^2 - p^2) \nonumber \\
&+& 2 p_1^{\mu}p^{\nu} (p\cdot p_1)),
\end{eqnarray}
\begin{eqnarray}
N_{18a}^{\mu\nu} &=& \frac{2 m_{\nu}^2 g^{\mu\nu}(p\cdot p_1)}{ (D-2)^2\xi^2}(2(p^2 +p_1^2)((D ((D -5)D +8) -20)\xi^2 +8 (D -2)\xi -(D -2)^2) \nonumber \\
&-& (2(((D -5)D +8)D +4)\xi^2 -8(D -2)\xi +(D -2)^2 ) m_{\nu}^2 ),
\end{eqnarray}
\begin{eqnarray}
N_{18b}^{\mu\nu} &=& -\frac{4g^{\mu\nu}(p\cdot p_ 1)}{(D-2)^2\xi^2}(p^2p_1^2((D ((D -5)D +8) -20)\xi^2 +8(D -2)\xi -(D -2)^2) \nonumber \\
&+& (D -4\xi -2)^2(p\cdot p_ 1)^2),
\end{eqnarray}
\begin{eqnarray}
N_{18c}^{\mu\nu} &=& -\frac{4(D -4\xi -2)^2}{(D-2)^2\xi^2} (p\cdot p_ 1)(p_1^{\mu} p_1^{\nu}(p^2 -m_{\nu}^2) +p^{\mu}p^{\nu}(p_1^2 -m_{\nu}^2 )),
\end{eqnarray}
\begin{eqnarray}
N_{18d}^{\mu\nu} &=& \frac{2(p_1^{\mu}p^{\nu} +p^{\mu}p_1^{\nu})}{(D-2)^2\xi^2}((2 (((D - 5) D + 8) D +4)\xi^2 -8(D -2)\xi +(D -2)^2 ) m_{\nu}^4 \nonumber \\
&-& 2(D -2)^2(D -1)\xi^2 (p^2 +p_1^2) m_{\nu}^2 +2(D -2)^2 (D -1)\xi^2 p^2 p_1^2 \nonumber \\
&+& 2(D -4\xi -2)^2 (p\cdot p_1 )^2 ).
\end{eqnarray}
As in the previous subsection, we have decomposed $N_{4}^{\mu\nu}$, $N_{7}^{\mu\nu}$, and $N_{18}^{\mu\nu}$ for clarity.

In what follows, we perform the shift $p_{\mu} \to p_{\mu} - k_{\mu}(1 - x)$, after which Eq.~\eqref{Pi1a} can be rewritten as
\begin{eqnarray}
    \Pi_1^{\mu\nu}=\Pi_{1a}^{\mu\nu}+\Pi_{1b}^{\mu\nu}+\Pi_{1c}^{\mu\nu}+\Pi_{1d}^{\mu\nu}+\Pi_{1e}^{\mu\nu}+\Pi_{1f}^{\mu\nu}+\Pi_{1g}^{\mu\nu},
\end{eqnarray} with
\begin{eqnarray}
\Pi_{1a}^{\mu\nu} &=& -ie^2\int_0^1dx\ \mu^{4-D}\int\frac{d^Dp}{(2\pi)^D}\frac{1}{(p^2-M^2)^4} \frac{8 g^{\mu \nu}}{(D-2)^2 D(D+2)\xi^2} ((k^2 (x-1)x) k^2 m_{\nu}^2 \nonumber \\
   &\times&  (-16 (D^2 -4)\xi +\xi^2 (8 (D-2)^2 (D^2-1)x^2 -8(D-2)^2 (D^2 -1)x+ D^4 -3D^3 \nonumber \\
   &-& 2D^2 +44D +56) +2(D+2)(D-2)^2 )+(D-2)k^4 (x-1)x (8 (D^2 -4)\xi +\xi^2 (D^4 \nonumber \\
   &-& 5D^2 +4(D-2) (D-1)(D+1)(D+3)x^2 -4(D-2)(D-1)(D+1)(D+3)x \nonumber \\
   &-& 16 D-28) -(D+2)(D-2)^2) -4(D-2)^2 (D-1)\xi^2 m_{\nu}^4),
\end{eqnarray}
\begin{eqnarray}
\Pi_{1b}^{\mu\nu} &=& ie^2\int_0^1dx\ \mu^{4-D}\int\frac{d^Dp}{(2\pi)^D}\frac{1}{(p^2-M^2)^4} \frac{8 k^{\mu}k^{\nu}}{(D-2)^2 D(D+2) \xi^2}(((x-1)x) 2 k^2 m_{\nu }^2 (-8 \nonumber \\
   &\times& (D^2 -4)\xi +2 \xi^2 (D^4-3 D^3-2 D^2+(D-2)^2 (D-1)(5 D+14)x^2 -(D-2)^2 \nonumber \\
   &\times& (D-1) (5 D +14)x +20D +8)+(D+2)(D-2)^2)+(D-2) k^4 (x-1)x    \nonumber \\
   &\times& (8(D^2-4)\xi + \xi^2 (D^4 +3D^3 -8D^2 +4(D-2)(D-1)(D+3)(D+4)x^2 \nonumber \\
   &-& 4(D-2)(D-1) (D+3)(D+4)x -28D -16)-(D+2) (D-2)^2) \nonumber\\
   &+& 8 (D-2)^2 (D-1) \xi^2 m_{\nu}^4),
\end{eqnarray}
\begin{eqnarray}
\Pi_{1c}^{\mu\nu} &=& -ie^2\int_0^1dx\ \mu^{4-D}\int\frac{d^Dp}{(2\pi)^D}\frac{1}{(p^2-M^2)^3} \frac{4 g^{\mu\nu}}{(D-2)^2 D(D+2)\xi^2} k^2((D-2)k^2 (x -1)x \nonumber \\
   &\times&  (24 (D^2-4)\xi +\xi^2 (4 (D-2)(D-1)(3D +2)(D+3)x^2 -4D((3 D-7) (D+3) \nonumber \\
   &\times&  D +4)x +(D (D+2)(3D -7) -44)D -8(6x +11)) -3(D+2)(D-2)^2) \nonumber \\
   &+& 2 m_{\nu}^2 (-8(D^2 -4)\xi +2 \xi^2 ((D-1)(D-2)^3 x^2 -(D-1)(D-2)^3 x +8(D+2)) \nonumber \\
   &+& (D+2)(D-2)^2)),
\end{eqnarray}
\begin{eqnarray}
\Pi_{1d}^{\mu\nu} &=& ie^2\int_0^1dx\ \mu^{4-D}\int\frac{d^Dp}{(2\pi)^D}\frac{1}{(p^2-M^2)^3} \frac{4 k^{\mu} k^{\nu}}{(D-2)^2 D(D+2)\xi^2} ((D-2) k^2 (x-1)x (24 \nonumber \\
   &\times&  (D^2-4)\xi +2 \xi^2 (D^4 +5D^3 -10D^2 +4(D-2)(D-1)(D+3)(D+6)x^2 -4(D \nonumber \\ &-& 2)(D-1)(D+3)(D+6)x -44D -24) -3(D+2) (D-2)^2) +2m_{\nu}^2 (-8(D^2 -4)\xi \nonumber \\
   &+& 2\xi^2 (D^4 -3D^3 -2D^2 +4(D-2)^2 (D-1)(D+4)x^2 -4(D-2)^2 (D-1)(D+4)x \nonumber \\
   &+& 20D +8) +(D+2)(D-2)^2) ),
\end{eqnarray}\begin{eqnarray}
\Pi_{1e}^{\mu\nu} &=& -ie^2\int_0^1dx\ \mu^{4-D}\int\frac{d^Dp}{(2\pi)^D}\frac{1}{(p^2-M^2)^2}\frac{4 g^{\mu \nu}}{D(D^2 -4)\xi^2} ( k^2 (8 (D^2-4)\xi +\xi^2 (-2 (3D \nonumber \\
   &-& 5) (D^2+D-4) Dx +D(D((D-1)D-4)-12) +2(D-2) (D-1) (D (3 D+7) \nonumber \\
   &-& 2) x^2 +8 (x-4)) -(D+2)(D-2)^2) -2(D-2)(D-1) (D+2)\xi^2 m_{\nu}^2 ),
\end{eqnarray}
\begin{eqnarray}
\Pi_{1f}^{\mu\nu} &=& ie^2\int_0^1dx\ \mu^{4-D}\int\frac{d^Dp}{(2\pi)^D}\frac{1}{(p^2-M^2)^2} \frac{4 k^{\mu} k^{\nu}}{D(D^2-4) \xi^2}(8 (D^2-4)\xi +2 \xi^2 ((D-2)\nonumber \\
   &\times&  (D-1)(D+4)(D+6)x^2-(D-2)(D-1)(D+4)(D+6)x +2((D-3)\nonumber \\
   &\times&  D-2) (D+2))-(D+2) (D-2)^2),
\end{eqnarray}
\begin{eqnarray}
\Pi_{1g}^{\mu\nu} &=& -ie^2\int_0^1dx\ \mu^{4-D}\int\frac{d^Dp}{(2\pi)^D}\frac{1}{(p^2-M^2)} \frac{4(D-2)(D-1)g^{\mu  \nu }}{D}, 
\end{eqnarray}
where now $M^2=k^2(x-1)x+m_\nu^2$. Next, once the momentum integrals have been evaluated, the full expression takes the form 
\begin{eqnarray}
\Pi_{1}^{\mu\nu} &=& -e^2\int_0^1dx\ \frac {2^{-D -1} e^{-\frac {1} {2} i\pi D}\pi^{-\frac {D} {2}} e^2\mu^4\left(\mu^2 \right)^{-\frac {D} {2}}\Gamma\left(1 - \frac {D} {2} \right)(-M)^{\frac {D - 8} {2}}}{3 (D - 2)\xi^2}(k^2 g^{\mu \nu} -k^{\mu} k^{\nu}) (D -2) \nonumber \\
   && \times (D -1)k^4 (x -1)^2 x^2(\xi^2(D(D^2 + 4 (D - 2) (D + 1) x^2 -4(D -2)(D +1)x -2D  \nonumber \\
   && -4) -8) +8(D -2)\xi -(D -2)^2) +k^2(x -1)x m_{\nu}^2(2\xi^2 (5 D^4 - 33 D^3 + 80 D^2 +2\nonumber \\
   && \times (11D -4)(D -2)^2 (D -1)x^2 -2(D -2)^2 (D-1)(11D -4)x -140D +160) +8 \nonumber \\
   && \times (7D -16)(D -2)\xi -((7D -16) (D -2)^2)) +2m_{\nu}^4(24(D -2)\xi +2\xi^2(20(D -1) \nonumber \\
   && \times  (D -2)^2 x^2 -20(D -1) (D -2)^2x +3((D -5)D +8)D -36) -3(D -2)^2).
\end{eqnarray}
Finally, expanding the above expression around $D = 4$ and integrating it over $x$ yields
\begin{eqnarray}\label{Pi1ml1}
\Pi_1^{\mu\nu} &=& -\frac{e^2 }{4 \pi^2 \epsilon' \xi^2}(6 \xi^2 -4 \xi +1)(k^2 g^{\mu \nu} -k^{\mu } k^{\nu }) +\frac{e^2}{4\pi^{2}\xi^{2}}(k^2 g^{\mu \nu} -k^{\mu } k^{\nu }) \nonumber\\
&&\times\Bigg[\frac{(-k^2(20 \xi ^2-12 \xi +1) m_{\nu}^2+k^4 (5 \xi -2) \xi -16 \xi^2 m_{\nu}^4)}{k^2 (4 m_{\nu }^2-k^2)} + \cot^{-1}(\sqrt{4 m_{\nu}^2 k^{-2}-1}) \nonumber \\ 
&& \times\frac{(-6 k^4 (6 \xi^2 -4 \xi +1) m_{\nu}^2 +4 k^2 (1 -2\xi)^2 m_{\nu}^4 +k^6 (6 \xi^2 -4\xi +1) +64 \xi^2 m_{\nu}^6) }{(k^2 (4 m_{\nu }^2-k^2))^{3/2}}\Bigg],\nonumber\\
\end{eqnarray}
where, again, $\frac{1}{\epsilon'} = \frac{1}{\epsilon} - \ln \frac{m}{\mu'}$, with $\epsilon = 4 - D$ and $\mu'^2 = 4\pi \mu^2 e^{-\gamma}$. As in the massive case, this expression is gauge invariant. However, the limit $m_\nu \to 0$ is not smooth. Taking into account the usual choice $\xi = \frac{1}{2}$ (see Ref.~\cite{vanNieuwenhuizen:1976bg}), we obtain the compact result
\begin{eqnarray}\label{Pi1ml2}
\Pi_1^{\mu\nu} &=& -\frac{e^2 }{2 \pi^2 \epsilon'}(k^2 g^{\mu \nu} - k^{\mu} k^{\nu}) 
- \frac{e^2}{4\pi^{2}}[(k^2+4m_\nu^2)k^{-2} - (2k^2+4m_\nu^2)k^{-2}\sqrt{k^2(4m_\nu^2-k^2)} \nonumber \\
&& \times \cot^{-1}(\sqrt{4 m_{\nu}^2 k^{-2}-1})] (k^2 g^{\mu \nu} - k^{\mu} k^{\nu}).
\end{eqnarray}
As we can see, the divergent term obtained above resembles that found in QED, suggesting that the massless case may be renormalizable.

\section{Conclusions}\label{CO}

We have calculated the photon self-energy arising from one-loop vacuum polarization in the RS theory within the VSR framework. We have considered both massive and massless spin-$3/2$ fields coupled to the Maxwell field. For the massive case, as expected, we obtain the gauge-invariant expression~\eqref{Pi1d}, which has a smooth limit as $m_\nu \to 0$ and allows us to recover the standard result~~\cite{Barua:1978ck}. Moreover, we find that the $m_\nu$-dependent divergent contribution is present only in the transverse projector ($k^2g^{\mu\nu}-k^\mu k^\nu$), as in the usual QED. For the massless case, we again obtain a gauge-invariant result~\eqref{Pi1ml1}, but this time it does not exhibit a smooth limit when $m_\nu \to 0$. Using the usual gauge choice $\xi = \frac{1}{2}$, which removes the last term in the propagator~\eqref{Gms}, we arrive at the compact expression~\eqref{Pi1ml2}, whose divergent term closely resembles that in QED. This similarity suggests that the massless theory may be renormalizable, a point that will be investigated in a forthcoming work.

\section*{Acknowledgments}

\hspace{0.5cm} MCA would like to thank FUNCAP, Funda\c{c}\~{a}o Cearense de Apoio ao Desenvolvimento Cient\'{i}fico e Tecnol\'{o}gico (Process No. DC3-0235-00076.01.00/24), and CNPq, Conselho Nacional de Desenvolvimento Cient\'{i}fico  e Tecnol\'{o}gico - Brasil (Process No. 304145/2025-4) for financial support. JF thanks FUNCAP (Project No.~PRONEM PNE0112-00085.01.00/16) and CNPq (Project No.~304485/2023-3) for financial support. JGL acknowledges CAPES (Finance Code 001) for financial support. TM thanks FAPEAL (Project No.~E:60030.0000002341/2022) and CNPq (Project No.~309360/2025-0) for financial support.

\global\long\def\link#1#2{\href{http://eudml.org/#1}{#2}}
\global\long\def\doi#1#2{\href{http://dx.doi.org/#1}{#2}}
\global\long\def\arXiv#1#2{\href{http://arxiv.org/abs/#1}{arXiv:#1 [#2]}}
\global\long\def\arXivOld#1{\href{http://arxiv.org/abs/#1}{arXiv:#1}}

%%%%%%%%%%%%%%%%%%%%%%%%%%%%%%%%%%%%%%%%%%%%%%%%%%%%%%%%%%%%%%%%%%%%%%%%%%%%


\begin{thebibliography}{99}

%\cite{Kostelecky:1988zi}
\bibitem{Kostelecky:1988zi}
V.~A.~Kostelecky and S.~Samuel,
%``Spontaneous Breaking of Lorentz Symmetry in String Theory,''
Phys. Rev. D \textbf{39}, 683 (1989).
%doi:10.1103/PhysRevD.39.683
%1692 citations counted in INSPIRE as of 02 Dec 2025

%\cite{Carroll:1989vb}
\bibitem{Carroll:1989vb}
S.~M.~Carroll, G.~B.~Field and R.~Jackiw,
%``Limits on a Lorentz and Parity Violating Modification of Electrodynamics,''
Phys. Rev. D \textbf{41}, 1231 (1990).
%doi:10.1103/PhysRevD.41.1231
%1454 citations counted in INSPIRE as of 02 Dec 2025

%\cite{Colladay:1996iz}
\bibitem{Colladay:1996iz}
D.~Colladay and V.~A.~Kostelecky,
%``CPT violation and the standard model,''
Phys. Rev. D \textbf{55}, 6760-6774 (1997)
%doi:10.1103/PhysRevD.55.6760
[arXiv:hep-ph/9703464 [hep-ph]].
%2107 citations counted in INSPIRE as of 02 Dec 2025

%\cite{Coleman:1998ti}
\bibitem{Coleman:1998ti}
S.~R.~Coleman and S.~L.~Glashow,
%``High-energy tests of Lorentz invariance,''
Phys. Rev. D \textbf{59}, 116008 (1999)
%doi:10.1103/PhysRevD.59.116008
[arXiv:hep-ph/9812418 [hep-ph]].
%1390 citations counted in INSPIRE as of 02 Dec 2025

%\cite{Colladay:1998fq}
\bibitem{Colladay:1998fq}
D.~Colladay and V.~A.~Kostelecky,
%``Lorentz violating extension of the standard model,''
Phys. Rev. D \textbf{58}, 116002 (1998)
%doi:10.1103/PhysRevD.58.116002
[arXiv:hep-ph/9809521 [hep-ph]].
%2728 citations counted in INSPIRE as of 02 Dec 2025

%\cite{Stecker:2001vb}
\bibitem{Stecker:2001vb}
F.~W.~Stecker and S.~L.~Glashow,
%``New tests of Lorentz invariance following from observations of the highest energy cosmic gamma-rays,''
Astropart. Phys. \textbf{16}, 97-99 (2001)
%doi:10.1016/S0927-6505(01)00137-2
[arXiv:astro-ph/0102226 [astro-ph]].
%227 citations counted in INSPIRE as of 02 Dec 2025

%\cite{Carroll:2001ws}
\bibitem{Carroll:2001ws}
S.~M.~Carroll, J.~A.~Harvey, V.~A.~Kostelecky, C.~D.~Lane and T.~Okamoto,
%``Noncommutative field theory and Lorentz violation,''
Phys. Rev. Lett. \textbf{87}, 141601 (2001)
%doi:10.1103/PhysRevLett.87.141601
[arXiv:hep-th/0105082 [hep-th]].
%1060 citations counted in INSPIRE as of 02 Dec 2025

%\cite{Kostelecky:2003fs}
\bibitem{Kostelecky:2003fs}
V.~A.~Kostelecky,
%``Gravity, Lorentz violation, and the standard model,''
Phys. Rev. D \textbf{69}, 105009 (2004)
%doi:10.1103/PhysRevD.69.105009
[arXiv:hep-th/0312310 [hep-th]].
%1547 citations counted in INSPIRE as of 02 Dec 2025

%\cite{Kostelecky:2006ta}
\bibitem{Kostelecky:2006ta}
V.~A.~Kostelecky and M.~Mewes,
%``Sensitive polarimetric search for relativity violations in gamma-ray bursts,''
Phys. Rev. Lett. \textbf{97}, 140401 (2006)
%doi:10.1103/PhysRevLett.97.140401
[arXiv:hep-ph/0607084 [hep-ph]].
%231 citations counted in INSPIRE as of 02 Dec 2025

%\cite{Kostelecky:2007zz}
\bibitem{Kostelecky:2007zz}
V.~A.~Kostelecky and M.~Mewes,
%``Lorentz-violating electrodynamics and the cosmic microwave background,''
Phys. Rev. Lett. \textbf{99}, 011601 (2007)
%doi:10.1103/PhysRevLett.99.011601
[arXiv:astro-ph/0702379 [astro-ph]].
%204 citations counted in INSPIRE as of 02 Dec 2025

%\cite{Kostelecky:1991ak}
\bibitem{Kostelecky:1991ak}
V.~A.~Kostelecky and R.~Potting,
%``CPT and strings,''
Nucl. Phys. B \textbf{359}, 545-570 (1991).
%doi:10.1016/0550-3213(91)90071-5
%790 citations counted in INSPIRE as of 02 Dec 2025

%\cite{Gambini:1998it}
\bibitem{Gambini:1998it}
R.~Gambini and J.~Pullin,
%``Nonstandard optics from quantum space-time,''
Phys. Rev. D \textbf{59}, 124021 (1999)
%doi:10.1103/PhysRevD.59.124021
[arXiv:gr-qc/9809038 [gr-qc]].
%1004 citations counted in INSPIRE as of 02 Dec 2025

%\cite{Bojowald:2004bb}
\bibitem{Bojowald:2004bb}
M.~Bojowald, H.~A.~Morales-Tecotl and H.~Sahlmann,
%``On loop quantum gravity phenomenology and the issue of Lorentz invariance,''
Phys. Rev. D \textbf{71}, 084012 (2005)
%doi:10.1103/PhysRevD.71.084012
[arXiv:gr-qc/0411101 [gr-qc]].
%93 citations counted in INSPIRE as of 02 Dec 2025

%\cite{Horava:2009uw}
\bibitem{Horava:2009uw}
P.~Horava,
%``Quantum Gravity at a Lifshitz Point,''
Phys. Rev. D \textbf{79}, 084008 (2009)
%doi:10.1103/PhysRevD.79.084008
[arXiv:0901.3775 [hep-th]].
%2716 citations counted in INSPIRE as of 02 Dec 2025

%\cite{Cognola:2016gjy}
\bibitem{Cognola:2016gjy}
G.~Cognola, R.~Myrzakulov, L.~Sebastiani, S.~Vagnozzi and S.~Zerbini,
%``Covariant Ho{\v{r}}ava-like and mimetic Horndeski gravity: cosmological solutions and perturbations,''
Class. Quant. Grav. \textbf{33}, no.22, 225014 (2016)
%doi:10.1088/0264-9381/33/22/225014
[arXiv:1601.00102 [gr-qc]].
%125 citations counted in INSPIRE as of 02 Dec 2025

%\cite{Coleman:1997xq}
\bibitem{Coleman:1997xq}
S.~R.~Coleman and S.~L.~Glashow,
%``Cosmic ray and neutrino tests of special relativity,''
Phys. Lett. B \textbf{405}, 249-252 (1997)
%doi:10.1016/S0370-2693(97)00638-2
[arXiv:hep-ph/9703240 [hep-ph]].
%532 citations counted in INSPIRE as of 02 Dec 2025

%\cite{Myers:2003fd}
\bibitem{Myers:2003fd}
R.~C.~Myers and M.~Pospelov,
%``Ultraviolet modifications of dispersion relations in effective field theory,''
Phys. Rev. Lett. \textbf{90}, 211601 (2003)
%doi:10.1103/PhysRevLett.90.211601
[arXiv:hep-ph/0301124 [hep-ph]].
%552 citations counted in INSPIRE as of 02 Dec 2025

%\cite{Andrianov:2001zj}
\bibitem{Andrianov:2001zj}
A.~A.~Andrianov, P.~Giacconi and R.~Soldati,
%``Lorentz and CPT violations from Chern-Simons modifications of QED,''
JHEP \textbf{02}, 030 (2002)
%doi:10.1088/1126-6708/2002/02/030
[arXiv:hep-th/0110279 [hep-th]].
%160 citations counted in INSPIRE as of 02 Dec 2025

%\cite{Alfaro:2006dd}
\bibitem{Alfaro:2006dd}
J.~Alfaro, A.~A.~Andrianov, M.~Cambiaso, P.~Giacconi and R.~Soldati,
%``On the consistency of Lorentz invariance violation in QED induced by fermions in constant axial-vector background,''
Phys. Lett. B \textbf{639}, 586-590 (2006)
%doi:10.1016/j.physletb.2006.06.075
[arXiv:hep-th/0604164 [hep-th]].
%70 citations counted in INSPIRE as of 02 Dec 2025

%\cite{Alfaro:2009mr}
\bibitem{Alfaro:2009mr}
J.~Alfaro, A.~A.~Andrianov, M.~Cambiaso, P.~Giacconi and R.~Soldati,
%``Bare and Induced Lorentz and CPT Invariance Violations in QED,''
Int. J. Mod. Phys. A \textbf{25}, 3271-3306 (2010)
%doi:10.1142/S0217751X10049293
[arXiv:0904.3557 [hep-th]].
%105 citations counted in INSPIRE as of 02 Dec 2025

%\cite{Cohen:2006ky}
\bibitem{Cohen:2006ky}
A.~G.~Cohen and S.~L.~Glashow,
%``Very special relativity,''
Phys. Rev. Lett. \textbf{97}, 021601 (2006)
%doi:10.1103/PhysRevLett.97.021601
[arXiv:hep-ph/0601236 [hep-ph]].
%358 citations counted in INSPIRE as of 02 Dec 2025

%\cite{Alfaro:2023qib}
\bibitem{Alfaro:2023qib}
J.~Alfaro,
%``Renormalization of Very Special Relativity gauge theories,''
JHEP \textbf{06}, 003 (2023)
%doi:10.1007/JHEP06(2023)003
[arXiv:2304.12933 [hep-ph]].
%6 citations counted in INSPIRE as of 03 Dec 2025

%\cite{Alfaro:2025uby}
\bibitem{Alfaro:2025uby}
J.~Alfaro,
%``Very Special Relativity Models: Infrared Regularization and Loop Corrections,''
Axioms \textbf{14}, no.6, 441 (2025).
%doi:10.3390/axioms14060441
%0 citations counted in INSPIRE as of 03 Dec 2025

%\cite{Dunn:2006xk}
\bibitem{Dunn:2006xk}
A.~Dunn and T.~Mehen,
``Implications of SU(2)(L) x U(1) symmetry for SIM(2) invariant neutrino masses,''
[arXiv:hep-ph/0610202 [hep-ph]].
%30 citations counted in INSPIRE as of 03 Dec 2025

%\cite{Cohen:2006ir}
\bibitem{Cohen:2006ir}
A.~G.~Cohen and S.~L.~Glashow,
``A Lorentz-Violating Origin of Neutrino Mass?,''
[arXiv:hep-ph/0605036 [hep-ph]].
%113 citations counted in INSPIRE as of 21 Dec 2025

%\cite{Cheon:2009zx}
\bibitem{Cheon:2009zx}
S.~Cheon, C.~Lee and S.~J.~Lee,
%``SIM(2)-invariant Modifications of Electrodynamic Theory,''
Phys. Lett. B \textbf{679}, 73-76 (2009)
%doi:10.1016/j.physletb.2009.07.007
[arXiv:0904.2065 [hep-th]].
%53 citations counted in INSPIRE as of 03 Dec 2025

%\cite{Alfaro:2015fha}
\bibitem{Alfaro:2015fha}
J.~Alfaro, P.~Gonz{\'a}lez and R.~{\'A}vila,
%``Electroweak standard model with very special relativity,''
Phys. Rev. D \textbf{91}, 105007 (2015)
%doi:10.1103/PhysRevD.91.129904
[arXiv:1504.04222 [hep-ph]].
%57 citations counted in INSPIRE as of 21 Dec 2025

%\cite{Alfaro:2013uva}
\bibitem{Alfaro:2013uva}
J.~Alfaro and V.~O.~Rivelles,
%``Non Abelian Fields in Very Special Relativity,''
Phys. Rev. D \textbf{88}, 085023 (2013)
%doi:10.1103/PhysRevD.88.085023
[arXiv:1305.1577 [hep-th]].
%50 citations counted in INSPIRE as of 03 Dec 2025

%\cite{Schwartz:2014sze}
\bibitem{Schwartz:2014sze}
M.~D.~Schwartz,
``Quantum Field Theory and the Standard Model,''
Cambridge University Press, 2014.
%ISBN 978-1-107-03473-0, 978-1-107-03473-0
%231 citations counted in INSPIRE as of 03 Dec 2025

%\cite{Alfaro:2019koq}
\bibitem{Alfaro:2019koq}
J.~Alfaro and A.~Soto,
%``Photon mass in very special relativity,''
Phys. Rev. D \textbf{100}, no.5, 055029 (2019)
%doi:10.1103/PhysRevD.100.055029
[arXiv:1901.08011 [hep-th]].
%31 citations counted in INSPIRE as of 03 Dec 2025

%\cite{Cohen:2006sc}
\bibitem{Cohen:2006sc}
A.~G.~Cohen and D.~Z.~Freedman,
%``SIM(2) and SUSY,''
JHEP \textbf{07}, 039 (2007)
%doi:10.1088/1126-6708/2007/07/039
[arXiv:hep-th/0605172 [hep-th]].
%56 citations counted in INSPIRE as of 21 Dec 2025

%\cite{Vohanka:2011aa}
\bibitem{Vohanka:2011aa}
J.~Vohanka,
%``Gauge Theory and SIM(2) Superspace,''
Phys. Rev. D \textbf{85}, 105009 (2012)
%doi:10.1103/PhysRevD.85.105009
[arXiv:1112.1797 [hep-th]].
%38 citations counted in INSPIRE as of 03 Dec 2025

%\cite{Gibbons:2007iu}
\bibitem{Gibbons:2007iu}
G.~W.~Gibbons, J.~Gomis and C.~N.~Pope,
%``General very special relativity is Finsler geometry,''
Phys. Rev. D \textbf{76}, 081701 (2007)
%doi:10.1103/PhysRevD.76.081701
[arXiv:0707.2174 [hep-th]].
%243 citations counted in INSPIRE as of 03 Dec 2025

%\cite{Muck:2008bd}
\bibitem{Muck:2008bd}
W.~Muck,
%``Very Special Relativity in Curved Space-Times,''
Phys. Lett. B \textbf{670}, 95-98 (2008)
%doi:10.1016/j.physletb.2008.10.028
[arXiv:0806.0737 [hep-th]].
%33 citations counted in INSPIRE as of 03 Dec 2025

%\cite{Sheikh-Jabbari:2008ybm}
\bibitem{Sheikh-Jabbari:2008ybm}
M.~M.~Sheikh-Jabbari and A.~Tureanu,
%``Realization of Cohen-Glashow Very Special Relativity on Noncommutative Space-Time,''
Phys. Rev. Lett. \textbf{101}, 261601 (2008)
%doi:10.1103/PhysRevLett.101.261601
[arXiv:0806.3699 [hep-th]].
%69 citations counted in INSPIRE as of 03 Dec 2025

%\cite{Das:2010cn}
\bibitem{Das:2010cn}
S.~Das, S.~Ghosh and S.~Mignemi,
%``Noncommutative Spacetime in Very Special Relativity,''
Phys. Lett. A \textbf{375}, 3237-3242 (2011)
%doi:10.1016/j.physleta.2011.07.024
[arXiv:1004.5356 [hep-th]].
%26 citations counted in INSPIRE as of 03 Dec 2025

%\cite{Alvarez:2008uy}
\bibitem{Alvarez:2008uy}
E.~Alvarez and R.~Vidal,
%``Very Special (de Sitter) Relativity,''
Phys. Rev. D \textbf{77}, 127702 (2008)
%doi:10.1103/PhysRevD.77.127702
[arXiv:0803.1949 [hep-th]].
%31 citations counted in INSPIRE as of 03 Dec 2025

%\cite{Ahluwalia:2010zn}
\bibitem{Ahluwalia:2010zn}
D.~V.~Ahluwalia and S.~P.~Horvath,
%``Very special relativity as relativity of dark matter: The Elko connection,''
JHEP \textbf{11}, 078 (2010)
%doi:10.1007/JHEP11(2010)078
[arXiv:1008.0436 [hep-ph]].
%111 citations counted in INSPIRE as of 03 Dec 2025

%\cite{Chang:2013xwa}
\bibitem{Chang:2013xwa}
Z.~Chang, M.~H.~Li, X.~Li and S.~Wang,
%``Cosmological model with local symmetry of very special relativity and constraints on it from supernovae,''
Eur. Phys. J. C \textbf{73}, no.6, 2459 (2013)
%doi:10.1140/epjc/s10052-013-2459-x
[arXiv:1303.1593 [astro-ph.CO]].
%49 citations counted in INSPIRE as of 03 Dec 2025

%\cite{Fan:2006nd}
\bibitem{Fan:2006nd}
J.~Fan, W.~D.~Goldberger and W.~Skiba,
%``Spin dependent masses and Sim(2) symmetry,''
Phys. Lett. B \textbf{649}, 186-190 (2007)
%doi:10.1016/j.physletb.2007.03.055
[arXiv:hep-ph/0611049 [hep-ph]].
%21 citations counted in INSPIRE as of 03 Dec 2025

%\cite{Freedman:1976py}
\bibitem{Freedman:1976py}
D.~Z.~Freedman and P.~van Nieuwenhuizen,
%``Properties of Supergravity Theory,''
Phys. Rev. D \textbf{14}, 912 (1976).
%doi:10.1103/PhysRevD.14.912
%312 citations counted in INSPIRE as of 03 Dec 2025

%\cite{Das:1976ct}
\bibitem{Das:1976ct}
A.~K.~Das and D.~Z.~Freedman,
%``Gauge Quantization for Spin 3/2 Fields,''
Nucl. Phys. B \textbf{114}, 271-296 (1976).
%doi:10.1016/0550-3213(76)90589-7
%100 citations counted in INSPIRE as of 03 Dec 2025

%\cite{Gates:1983nr}
\bibitem{Gates:1983nr}
S.~J.~Gates, M.~T.~Grisaru, M.~Rocek and W.~Siegel,
%``Superspace Or One Thousand and One Lessons in Supersymmetry,''
Front. Phys. \textbf{58}, 1-548 (1983)
%1983,
%ISBN 978-0-8053-3161-5
[arXiv:hep-th/0108200 [hep-th]].
%963 citations counted in INSPIRE as of 21 Dec 2025

%\cite{Delgado-Acosta:2009ulg}
\bibitem{Delgado-Acosta:2009ulg}
E.~G.~Delgado-Acosta and M.~Napsuciale,
%``Compton scattering off elementary spin 3/2 particles,''
Phys. Rev. D \textbf{80}, 054002 (2009)
%doi:10.1103/PhysRevD.80.054002
[arXiv:0907.1124 [hep-th]].
%17 citations counted in INSPIRE as of 03 Dec 2025

%\cite{Antoniadis:2022jjy}
\bibitem{Antoniadis:2022jjy}
I.~Antoniadis, A.~Guillen and F.~Rondeau,
%``Massive gravitino scattering amplitudes and the unitarity cutoff of the new Fayet-Iliopoulos terms,''
JHEP \textbf{01}, 043 (2023)
%doi:10.1007/JHEP01(2023)043
[arXiv:2210.00817 [hep-th]].
%6 citations counted in INSPIRE as of 03 Dec 2025

%\cite{Araujo:2024bug}
\bibitem{Araujo:2024bug}
M.~C.~Ara{\'u}jo, J.~G.~Lima, J.~Furtado and T.~Mariz,
%``Bhabha-like scattering in the Rarita{\textendash}Schwinger model at finite temperature,''
Eur. Phys. J. C \textbf{85}, no.3, 314 (2025)
%doi:10.1140/epjc/s10052-025-14002-6
[arXiv:2412.19910 [hep-th]].
%0 citations counted in INSPIRE as of 03 Dec 2025

%\cite{deJong:1992wm}
\bibitem{deJong:1992wm}
F.~de Jong and R.~Malfliet,
%``Covariant description of the Delta in nuclear matter,''
Phys. Rev. C \textbf{46}, 2567-2581 (1992).
%doi:10.1103/PhysRevC.46.2567
%42 citations counted in INSPIRE as of 03 Dec 2025

%\cite{Pascalutsa:1999zz}
\bibitem{Pascalutsa:1999zz}
V.~Pascalutsa and R.~Timmermans,
%``Field theory of nucleon to higher spin baryon transitions,''
Phys. Rev. C \textbf{60}, 042201 (1999)
%doi:10.1103/PhysRevC.60.042201
[arXiv:nucl-th/9905065 [nucl-th]].
%177 citations counted in INSPIRE as of 03 Dec 2025

%\cite{Bernard:2003xf}
\bibitem{Bernard:2003xf}
V.~Bernard, T.~R.~Hemmert and U.~G.~Meissner,
%``Infrared regularization with spin 3/2 fields,''
Phys. Lett. B \textbf{565}, 137-145 (2003)
%doi:10.1016/S0370-2693(03)00538-0
[arXiv:hep-ph/0303198 [hep-ph]].
%77 citations counted in INSPIRE as of 03 Dec 2025

%\cite{Gomes:2022btc}
\bibitem{Gomes:2022btc}
M.~Gomes, T.~Mariz, J.~R.~Nascimento and A.~Y.~Petrov,
%``Lorentz-breaking Rarita-Schwinger model,''
Phys. Scripta \textbf{98}, no.12, 125260 (2023)
%doi:10.1088/1402-4896/ad0e9c
[arXiv:2211.12414 [hep-th]].
%6 citations counted in INSPIRE as of 03 Dec 2025

%\cite{Gomes:2023qkj}
\bibitem{Gomes:2023qkj}
M.~Gomes, J.~G.~Lima, T.~Mariz, J.~R.~Nascimento and A.~Y.~Petrov,
%``Non-Abelian Carroll{\textendash}Field{\textendash}Jackiw term in a Rarita-Schwinger model,''
Phys. Lett. B \textbf{845}, 138141 (2023)
%doi:10.1016/j.physletb.2023.138141
[arXiv:2308.16308 [hep-th]].
%4 citations counted in INSPIRE as of 03 Dec 2025

%\cite{Gomes:2024qya}
\bibitem{Gomes:2024qya}
M.~Gomes, J.~G.~Lima, T.~Mariz, J.~R.~Nascimento and A.~Y.~Petrov,
%``Carroll{\textendash}Field{\textendash}Jackiw term in a massless Rarita-Schwinger model,''
Phys. Lett. B \textbf{864}, 139408 (2025)
%doi:10.1016/j.physletb.2025.139408
[arXiv:2410.01550 [hep-th]].
%0 citations counted in INSPIRE as of 03 Dec 2025

%\cite{Mandelstam:1982cb}
\bibitem{Mandelstam:1982cb}
S.~Mandelstam,
%``Light Cone Superspace and the Ultraviolet Finiteness of the N=4 Model,''
Nucl. Phys. B \textbf{213}, 149-168 (1983).
%doi:10.1016/0550-3213(83)90179-7
%1014 citations counted in INSPIRE as of 03 Dec 2025

%\cite{Leibbrandt:1983pj}
\bibitem{Leibbrandt:1983pj}
G.~Leibbrandt,
%``The Light Cone Gauge in Yang-Mills Theory,''
Phys. Rev. D \textbf{29}, 1699 (1984).
%doi:10.1103/PhysRevD.29.1699
%383 citations counted in INSPIRE as of 03 Dec 2025

%\cite{Alfaro:2016pjw}
\bibitem{Alfaro:2016pjw}
J.~Alfaro,
%``Mandelstam-Leibbrandt prescription,''
Phys. Rev. D \textbf{93}, no.6, 065033 (2016)
[erratum: Phys. Rev. D \textbf{94}, no.4, 049901 (2016)]
%doi:10.1103/PhysRevD.94.049901
[arXiv:1603.06453 [hep-th]].

%\cite{Barua:1978ck}
\bibitem{Barua:1978ck}
D.~Barua and S.~N.~Gupta,
%``Electromagnetic Interaction of Higher Spin Fields,''
Phys. Rev. D \textbf{17}, 2028-2037 (1978).
%doi:10.1103/PhysRevD.17.2028
%4 citations counted in INSPIRE as of 21 Dec 2025

%\cite{vanNieuwenhuizen:1976bg}
\bibitem{vanNieuwenhuizen:1976bg}
P.~van Nieuwenhuizen and J.~A.~M.~Vermaseren,
%``One Loop Divergences in the Quantum Theory of Supergravity,''
Phys. Lett. B \textbf{65}, 263-266 (1976).
%doi:10.1016/0370-2693(76)90178-7
%63 citations counted in INSPIRE as of 21 Dec 2025

\end{thebibliography}
\end{document}